\documentclass[preprint]{iucr}              
\usepackage{graphicx}
\usepackage{amsmath}
\usepackage{xcolor}
\usepackage{soul}
\sethlcolor{yellow} 
\usepackage[utf8]{inputenc}
\usepackage[T1]{fontenc}
\DeclareUnicodeCharacter{03B1}{\ensuremath{\alpha}}
\DeclareUnicodeCharacter{2265}{\ensuremath{\ge}}
\DeclareUnicodeCharacter{2212}{\ensuremath{-}}
\DeclareUnicodeCharacter{202F}{\ensuremath{\!}}
\DeclareUnicodeCharacter{2062}{\enspace}
\DeclareUnicodeCharacter{202F}{\thinspace}        
\DeclareUnicodeCharacter{2062}{\textvisiblespace} 
\usepackage{textcomp}
     \journalcode{J}              
\sethlcolor{yellow} 
\begin{document}                  



\title{Structural analysis of co-sputtered Cu-Nb and Cu-Pd textured thin films}


\cauthor[a]{Claudia}{Cancellieri}{claudia.cancellieri@empa.ch}{}
\author[a]{Giacomo}{Lorenzin}
\author[b]{Yeliz}{Unutulmazsoy}
\author[b]{Andriy}{Lotnyk}
\author[c]{Daniel}{Ariosa}

\aff[a]{Empa, Swiss Federal Laboratories for Materials Science and Technology, Laboratories for Joining Technologies and Corrosion, \"{U}berlandstrasse 129, 8600 D\"{u}bendorf, Switzerland}
\aff[b]{Leibniz Institute of Surface Engineering (IOM) Permoserstrasse 15, 04318 Leipzig, Germany}
\aff[c]{Instituto de F\'{i}sica, Facultad de Ingenier\'{i}a, Universidad de la Rep\'{u}blica, Herrera y Reissig 565, C.C. 30, 11000 Montevideo, Uruguay}









\maketitle                        

\begin{synopsis}
This study investigates the structural evolution of co-sputtered Cu-based thin films with varying concentrations of Nb and Pd, using X-ray diffraction (XRD), scanning electron microscopy (SEM), and energy-dispersive X-ray spectroscopy (EDX). The introduction of Nb and Pd into the Cu matrix significantly impacts the films' crystallographic and microstructural properties, offering insights into the complex interplay between elemental concentration and thin film growth under non-equilibrium conditions. A random intercalation model is proposed to explain the observed structural trends, including peak shifting, broadening, and reduced coherence. The model accounts for interstitial and substitutional cases and is applicable to other co-deposited thin film systems.
\end{synopsis}

\begin{abstract}
Structural characterization of nanoscale-two-metal-phase systems, which exhibit partial, complete, or no mixing when co-sputtered with a few percent of a minority element, is extremely challenging. Co-sputtering two metals at room temperature results in frozen disorder within the deposited films. Distinguishing the contribution of each metal phase, determining the distribution and self-organization of the second constituent element within the lattice, accurately quantifying the extra element content, and assessing internal disorder through diffraction analysis are complex and require the development of a suitable model to fit diffraction patterns from various geometries. Here, we present a model to describe the structural distribution of alloy elements in magnetron-sputtered Cu thin films, exploring two contrasting cases: 1) with the mutually immiscible Nb and 2) with Pd, which has a negative heat of mixing with Cu, forming stable alloys. A comparison between X-ray diffraction data and energy-dispersive X-ray spectroscopy-derived elemental distribution is discussed.
\end{abstract}


\section{Introduction}

The introduction of extra elements in metals is highly desirable to tune material properties, for example, catalytic activity, electrical conductivity, hardness, oxidation resistance, wear resistance, reduction kinetics and thermal stability. The presence of another constituent element in the metallic lattice can affect the grain size, which is found to strongly affect the oxidation and reduction kinetics of Cu films at elevated temperatures \cite{UNUTULMAZSOY2022152896,CuoxidationYeliz}. Copper-based materials have been widely used for different applications due to their excellent electrical and thermal conductivity, good workability and costs  \cite{Lu2004422}. However, they suffer from poor mechanical 
properties, which can be solved by alloying Cu with extra elements like Nb \cite{DING2021108662}, Si \cite{LI2018129} or C \cite{tang2008friction}.
The presence of either insoluble particles or alloying elements can yield to higher mechanical strength in metallic systems although the efficiency of the final material strongly depends on how homogeneous is the distribution of the additional elements. 

Progress in modern processing techniques allows to artificially create a multitude of new alloys in systems that are immiscible in thermodynamic equilibrium \cite{MA2005413} by employing non-equilibrium processing to
overcome the effects of the positive entalphy of mixing, $\Delta$H, opening up new opportunities for tailoring materials properties. The actual distribution of the introduced elements in the metal matrix is paramount in determining the final properties and performance of materials. Especially highly immiscible alloys do not homogenize completely  and single-phase
alloys, such as amorphous phases, supersaturated crystalline solid solutions, and some metastable intermediate compounds, can be formed. Structural inhomogeneity characterization in such systems is challenging especially for low atomic concentrations and for perfect solid solutions. 

A widely studied example of immiscible binary system is Cu-Nb. Nb has a positive enthalpy of mixing with Cu \cite{CuNbphase}. Immiscible alloys are widely used as conducting materials for electrical applications, which require electrical conductivity and high mechanical strength simultaneously \cite{botcharova2006mechanical, LEI2013367}.   Cu–Nb is a model system for gaining insight into the mechanical properties of nanocomposites \cite{botcharova2006mechanical} and understanding the behavior of metallic interfaces under extreme environments. Cu–Nb alloy, that is completely immiscible, can form both a crystalline and an amorphous phase \cite{WANG2007157}. These studies provided key findings of the mechanical behavior of nanocrystalline and nanolayered metals produced by film deposition, as well as nanostructured alloys produced by severe plastic deformation. The ultimate goal is to make materials more reliable for a wide variety of applications (microelectronics, bonding, etc.). Adapting the thermal, electronic and mechanical properties of such structures requires precise knowledge of their structure at the nanometric scale which is not always easy to access. 
Thus, the properties of nanostructured materials always have to be associated with the grain size, which, at the same time, has a great influence on the volume fraction of the grain boundaries. With respect to the large number of grain boundaries, one can estimate that the mechanical strength of nanocrystalline metallic materials is enhanced as the dislocation motion is also hindered by grain boundaries.
Nb was found to increase the mechanical
strength of Cu alloys when added as high concentration solutes to form a
high density of nano-scale precipitates \cite{Banerjee}. In particular for Cu-Nb it was found an important effect of plastic deformation on the self organization of Nb precipitates and chemical mixing \cite{WANG2014276}. For ultra-thin films, the competition between phase separation and strain energy reduction enables the formation of self-assembled lateral multilayers of immiscible elements, as templates of modulated nanostructures \cite{PhysRevLett.88.186101}. 

Across the broad spectrum of alloys, the vast majority of them are formed in systems with a negative heat of mixing (the terms of heat of mixing and enthalpy of mixing are used interchangeably hereafter). In other words, the constituent elements
have the tendency to spontaneously alloy on atomic scale, due to the reduction of
Gibbs free energy upon intermixing. At ambient temperature, one observes a variety
of alloy phases such as solid solutions and intermetallic compounds. Their stability ranges depend on the thermodynamic properties of these competing phases as a function of alloy composition, and the kinetic parameters employed in processing. Pd-Cu bimetallic system is mainly used as catalyst \cite{DAI20119369}, which promotes the catalytic performance of HCOOH oxidation and enhance CO poisoning tolerance
compared to the pure Pd catalyst \cite{ZHANG2019730}. Cu-Pd alloys with near-equiatomic compositions are considered efficient catalysts and therefore are used for the fabrication of membranes for hydrogen purification \cite{OPALKA2007583}. The copper-based alloys with low palladium contents can be used
as electrical conductors that are characterized by adequate combination of high strength and corrosion
resistance \cite{volkov2016effect}. Contrarily to Nb, Pd is completely miscible with Cu \cite{de1988cohesion} forming also stable compounds and binary alloys. 

Both miscible and immiscible alloys present significant challenges in controlling and accessing nanostructure features which are crucial for applications and performances. Phase
distribution, self-organized composite morphologies, porosity, grain size, and interface
shape arise from the complex interplay of various dynamic mechanisms,
including deposition techniques and parameters, surface diffusion, grain coarsening, spinodal decomposition,
and others. Therefore, a deep understanding of thin film structure and the distribution
of alloying elements is crucial for achieving reliable and enhanced functionalities
in nanostructured thin films. In this work, we present an XRD model applied to both theta-2theta diffractograms and in-plane analysis using rocking curves (RC)
to investigate the effect of different contents of Nb and Pd on the Cu lattice. Comparison with scanning electron microscopy (SEM) and energy dispersive X-ray (EDX) planar and cross sectional analysis is also presented. This work is motivated by the future investigations that we plan to conduct on similar samples to understand the kinetics of oxidation and reduction of Cu in the presence of Nb and Pd. The present investigation will guide the optimization of co-sputtering parameters for future device fabrication and designing of Cu-based alloys with tailored properties for advanced applications in catalysis, microelectronics, and energy systems.

\section{Methods}

Cu-Nb and Cu-Pd thin films were grown on sapphire Al$_2$O$_3$ [0001] substrates at room temperature by DC magnetron sputtering. The Cu-Nb and Cu-Pd thin films of (100 $\pm$ 10) nm thickness are prepared by co-sputtering to obtain different nominal concentrations ranging from 4 to 10 at.\% for Nb and from 8 at.\%, to 39 at.\% for Pd. Pure Cu films, 100 $\pm$ 5 nm thick, are also grown for comparison. The growth rates have been measured prior to every deposition with a Bruker Dextat XTL profilometer. The deposition parameters are reported in Table 1.
\begin{table}
\caption{Deposition parameters used to obtain different Nb, Pd concentrations. The at.\% reported are obtained by XRF measurements.}
\begin{tabular}{lllcr}      
at\%.   & Cu Power (W)        & Nb power (W) & Pd Power (W)     & Ar Pressure (Pa)      \\
\hline
100 Cu    & 120      & -    & -     & 0.27  \\
96Cu-4Nb    & 140     & 20     & -     & 0.27  \\
94Cu-6Nb      & 120     & 20      & -   & 0.27  \\
92Cu-8Nb      & 140      & 30      & -   & 0.27 \\
88Cu-12Nb      & 120      & 40      & -   & 0.27 \\
61Cu-39Pd      & 140      & -     & 10   & 0.7 \\
89Cu-11Pd      &150     & -     & 10   & 0.7 \\
92Cu-8Pd      & 180      & -     & 10   & 0.7 \\
\end{tabular}
\end{table}
The power of Cu \footnote{The growth rate of Cu  was found to depend on the usage of the target. Samples produced in different period of times, if grown with the same magnetron power, showed a different growth rate.} has been adapted to have a fixed growth rate of $\sim$ 0.25 nm/sec. The minimum target power in order to stabilize a continuous plasma for Nb and Pd target was set to 20 and 10 W, respectively. 
The actual stoichiometry of co-sputtered samples is measured with a Fischer XDV-SDD X-ray Fluorescence (XRF) system equipped with a Rh X-ray source. The film structure and texture analysis have been derived by the XRD results reported in this paper.
A Bruker D8 Discover X-ray diffractometer has been used in Bragg Brentano geometry to measure coupled 2$\theta$ scans and Rocking curves (RC) using Cu K$\alpha$ radiation ($\lambda$= 1.541 \AA) at 40 kV and 40 mA. The diffractometer was equipped with a Lynxeye 1D detector, resolution of roughly 0.02° (full-width at half-maximum) and Goebel mirror mounted in the primary beam to collimate the beam for the coupled and omega scans. 
Pole figures were acquired using a circular slit of 1 mm in diameter to mimic point focus geometry. The angular tilt range of $\psi$ angle was from 0 to 80 $^\circ$, while in-plane $\phi$ angle was varied from 0 to 360$^\circ$. The $2\theta$ Bragg condition was fixed to the (111) reflection found by performing previously the coupled 2$\theta$ scan. 
Planar images were acquired in a FEI Helios NanoLab 660 dual beam scanning electron microscopy (SEM). The surfaces of the samples were imaged with an SEM instrument of the Hitachi S-4800 field emission gun using an acceleration voltage of 2 kV and with the detector in secondary electron (SE) mode. Elemental mapping is performed using a  Super-X energy dispersive X-ray (EDX) detector operating at 10 kV (Oxford Instruments X-MaxN 150 mm$^2$). Cross-sectional thin film lamellae were prepared from thin films using a focused ion beam (FIB) with a Zeiss Auriga Dual-Beam System. Transmission electron microscopy (TEM) imaging was performed on the lamellae using a probe Cs-corrected FEI Titan3 G2 60-300 microscope, operated at an accelerating voltage of 300 kV. FEI Super-X detector EDX system was used to acquire EDX maps in scanning TEM (STEM) mode. The SEM and STEM analyses, as well as the pole figures, were performed on selected samples that span the full range of Nb and Pd concentrations studied—covering low, intermediate, and high content—to capture the evolution of microstructure across the series.
\subsection{Modeling of XRD data}
We propose to adjust the experimental XRD data of  Cu/Nb and Cu/Pd co-sputtered films within the framework of a model for random intercalation of a guest phase (Nb or Pd) within the host (111) Cu matrix. This choice is justified by the low diffusivity of the guest species within the host matrix for samples condensed at room temperature from the vapor phase. The model proposed applies only to the crystalline phases present in the films and we assumed here that any amorphous regions, if present, are negligible. The diffraction model is the one proposed earlier to study the diffraction anomalies in superconducting YBCO films \cite{AriosaTsaneva2005}. It considers the position of the diffracting atoms of the host lattice as stochastically displaced from their periodic position due to the intercalation of the guest atoms. The value of each displacement is considered to be the same, so the crystal is subject to a frozen disorder that affects its diffraction pattern cumulatively. Such disorder reduces the phase correlation between the diffracted contributions modifying the regular periodic positions of peak maxima in momentum space.  Peak shapes are also affected becoming asymmetric and their intensities are reduced. The model has been successfully applied in other layered materials, essentially BSCCO films \cite{AriosaCancellieri2007, CancellieriLin2007, YelpoFavre2020}. 

In the present work, the diffraction pattern observed in ($\theta-2\theta$) scans can be written on the basis of the mean amplitude $A(q)$ averaged over the frozen disorder of intercalates, as shown in \cite{AriosaTsaneva2005}: 
\begin{equation}
    \bigl \langle A(q) \bigr \rangle= \bar f(q) \sum_{n=1}^{N} \tilde P(q)^{n}   
\end{equation}  
\begin{center}
    with
\end{center}
\begin{equation}
\bar f(q)= (1-\alpha)f_0(q) +\alpha f_1(q) \hspace{1cm} \text{and} \hspace{1cm} \tilde P(q)=[(1-\alpha) +\alpha e^{iq\delta}]e^{iqc_0}
\end{equation}  In equations (1) and (2), $q=\frac{4\pi\sin{\theta}}{\lambda}$ is the scattering wave vector, where $\lambda$ is the x-ray wavelength and $\theta$ half the scattering angle. The parameter $\alpha$ is the probability to find a host diffracting unit doped with a guest atom along the X-ray beam path,  which depends directly on the guest fraction $p$. The average structure factor $\bar f(q)$ is the resulting weighted average of $f_0(q)$ and $f_1(q)$, the local structure factors of the unperturbed diffracting unit and the one resulting from the guest intercalation. The grain size is given in terms of the number of diffracting planes $N$, $c_0$ is the out-of-plane regular distance between crystallographic planes of the host along the diffraction direction and $\delta$ is the local displacement caused by a single intercalate. In previous contributions, the emphasis was on fitting the non-monotonic shift of peak positions relative to the regular periodic structure. This was an easy task for highly oriented structures with unit cell parameters larger than 10 {\AA} along the diffraction direction, thus exhibiting many diffraction orders in the available angle range. Under these conditions, it was not necessary to take into account the intrinsic gaussian positional disorder of the host matrix or the particle size distribution, since adjusting for the many peak maxima deviations was sufficient to accurately extract the  probability $\alpha$ and the unit cell distortion $\delta$. On the contrary, in the current situation we are dealing with very few available diffraction orders, which forces us to focus on adjusting not only the position, but also the shape of the peak for a given order. To achieve this goal in a realistic way, intrinsic gaussian displacement disorder  and particle size distribution must be added to the model. The notation for the corresponding standard deviations are, respectively, $\sigma$ and $\Sigma$. The diffracted intensity for a  given grain size of N planes, with the amplitude in expression (1) corrected for gaussian displacement disorder reads:
\begin{equation}
I(q)=|\bigl \langle A(q) \bigr \rangle_{\sigma}|^2= |\bar f(q)|^2 \frac{1+|Z|^{2N}-2\Re(Z^N)}{|1-Z|^2} \quad\text{with}\quad Z(q)=e^{-\frac{(q\sigma)^2}{2}}\tilde P(q) 
\end{equation}
In the expression above, the symbol $\Re$ reads for the real part.  After performing on expression (3) the gaussian average over the grain size distribution, the final expression for the diffracted intensity is obtained:
 \begin{equation}
     \langle I(q)\rangle _\Sigma =|f(q)|^2 \frac{1+|Z|^{2(\bar N+\Sigma^2ln|Z|)}-2\Re(Z^{\bar N+\frac{1}{2}\Sigma^2lnZ})}{|1-Z|^2} 
\end{equation}
From the fit of the experimental XRD data one can extract the fraction $p(\alpha)$ of the guest atoms, the value $\delta$ of the fixed displacement associated with the intercalar defects, the average grain size $\bar N$(and its standard deviation $\Sigma$), as well as the standard deviation $\sigma$ characterizing the intrinsic crystal disorder of the host matrix.
Note that this approach goes beyond the granular solid-solution model. Indeed, our model relies on the stochastic nature of the position occupied by the guest atoms and the resulting detrimental effect on the crystal coherence. The shift of the peak position and its asymmetric shape are determined by expressions (1) to (4) without any further assumptions. \\
The above model applies to both, interstitial and substitutional random distribution of the secondary element. However, some model parameters must be specified for each of these two different situations.
\subsection*{\underline{Interstitial element inclusion}:}
In our specific case, the Cu matrix is a [111] textured film. Guest interstitial atoms are assumed to occupy random positions between two consecutive Cu(111) planes. More precisely, the guest atoms occupy an octahedral interstitial site at the center of the face-centered cubic (FCC) Cu unit cell. This assumption, would presumably apply more for Nb than for Pd, due to the immiscibility of the Cu-Nb system and by the fact that our films are grown at room temperature from a mixed vapor phase. Indeed, because of the very low mobility at this temperature, the guest atoms attach to the place where they ``fall" and influence the position of the copper structure that then develops above. Under these conditions, the different coordination of the two elements makes substitution unlikely.
\noindent The values of the local structure factors $f_0(q)$ and $f_1(q)$ for the interstitial guest/Cu samples are:
\begin{equation}
    f_0(q)=2f_{Cu}(q) \quad\text{and}\quad f_1(q)=2f_{Cu}(q)+f_{G}(q)e^{i\frac{ qc_1}{2}} ,
\end{equation}
where $f_{Cu}(q)$ and $f_{G}(q)$ are the atomic scattering factors for Cu and guest atoms, respectively. The guest atom is assumed to be in the center of a Cu FCC UC, between two consecutive planes (111). Thus, $f_0 (q)$ is just the sum of the structure factors of the Cu atoms lying on the (111) plane in an FCC UC.
The $f_1 (q)$ is the structure factor of the system formed by the Cu atoms on the (111) plane and the interstitial guest atom in the center of the cubic cell, trapped in between two consecutive (111) planes.
Notice that the phase reference is the (111) plane, then the guest is affected by the phase factor $e^{i\frac{ qc_1}{2}}$, $c_1$ being the locally modified (111) interplanar distance. The factor 2 affecting the Cu scattering factor in (5) is due to the double surface density of Cu compared to a single interstitial guest atom in the (111) projection. 
Accordingly, since the probability $\alpha$ is the number of perturbed diffraction units divided the total number of units, the relationship with the guest fraction is  $p\approx\alpha/2$ in the interstitial scenario.
The interplanar distance of Cu(111) is $c_0=a_{Cu}/\sqrt{3}$. The theoretical value of the cubic lattice constant of a bulk Cu is $a_{Cu}=3.615$ {\AA}. The measured values reported in the following sections range from 3.609 to 3.660 \AA.
\subsection*{\underline{Substitutional case}:}
In this case, substitution of Cu with the guest atom is considered. This scenario would presumably be adequate to the Cu/Pd system, since it presents an extremely good miscibility. The guest atom is assumed to randomly occupy positions corresponding to the Cu atom in the host matrix. We will assume that the local distortion caused by this substitution symmetrically affects the two (111) planes on each side of the doped plane. In order to use the intercalation model described in equations (1) to (4), it is then simpler to consider a double unit cell and adapt the structure factors accordingly. 

The changes are as follows:
\begin{gather}
    \tilde{c_0}=2 c_0 \qquad, 
    \qquad \tilde{c_1}=2 c_1= 2(c_0+\delta), \nonumber\\
    \tilde{f}_0(q)=f_{Cu}(q)(1+\cos{(qc_0)})\quad, 
    \quad \tilde{f}_1(q)=f_{G}(q)+f_{Cu}(q)\cos{(qc_1)},\\
   \tilde{f}(q)=(1-2\alpha)\tilde{f}_0(q)+2\alpha\tilde{f}_1(q) \qquad\text{and}\qquad\tilde{N}=N/2 \nonumber
\end{gather}
 The symbol ( $\Tilde{ }$ ) in equations (6) denotes the modified parameters. In $\tilde{f}_0(q)$, a Cu atom is considered at the center, while for $\tilde{f}_1(q)$, the Cu atom is substituted with a guest atom, both with zero phase. The model can now be used for substitutional element provided the variable changes shown above have been made. The relationship between the probability $\alpha$ and the guest fraction $p$ in the substitutional scenario is immediate: $p=\alpha$. This is because the guest itself is the perturbed unit while the Cu atoms in the matrix are the unperturbed ones.

\noindent In both cases, the residual strain undergone by the films must be taken into account. We define the parameter $\epsilon$ (in \%) which quantifies the strain along the [111] direction in the Cu matrix.

\noindent As we shall see in the following, the concepts of grain and grain size are not well defined in our context. Indeed, what we extract from XRD modeling is the crystal coherence length due to both, random intercalation and structural defects. However, we will use these terms as a shortcut.

\newpage

\section{Results and Discussion}
\subsection{Texture analysis by XRD}
\begin{figure}
\caption{Pole figures around Cu\{111\} for a) pure Cu, b) Cu with different Nb contents and c) Cu with different Pd concentrations.The white dashed line indicate the peaks at 62$^\circ$ tilt corresponding to the peaks of the (2 -1 -1 3)  sapphire substrate reflections. The balck dashed line correspond to the tilt angle of the \{111\} planes in FCC structure of the films. The color palette is the same for all the pole figures and the minimum and maximum intensity values are set individually for each sample. This choice preserves low-intensity features that would otherwise be lost if a single global scale were imposed}\label{pole}
\includegraphics[width=1.0\textwidth]{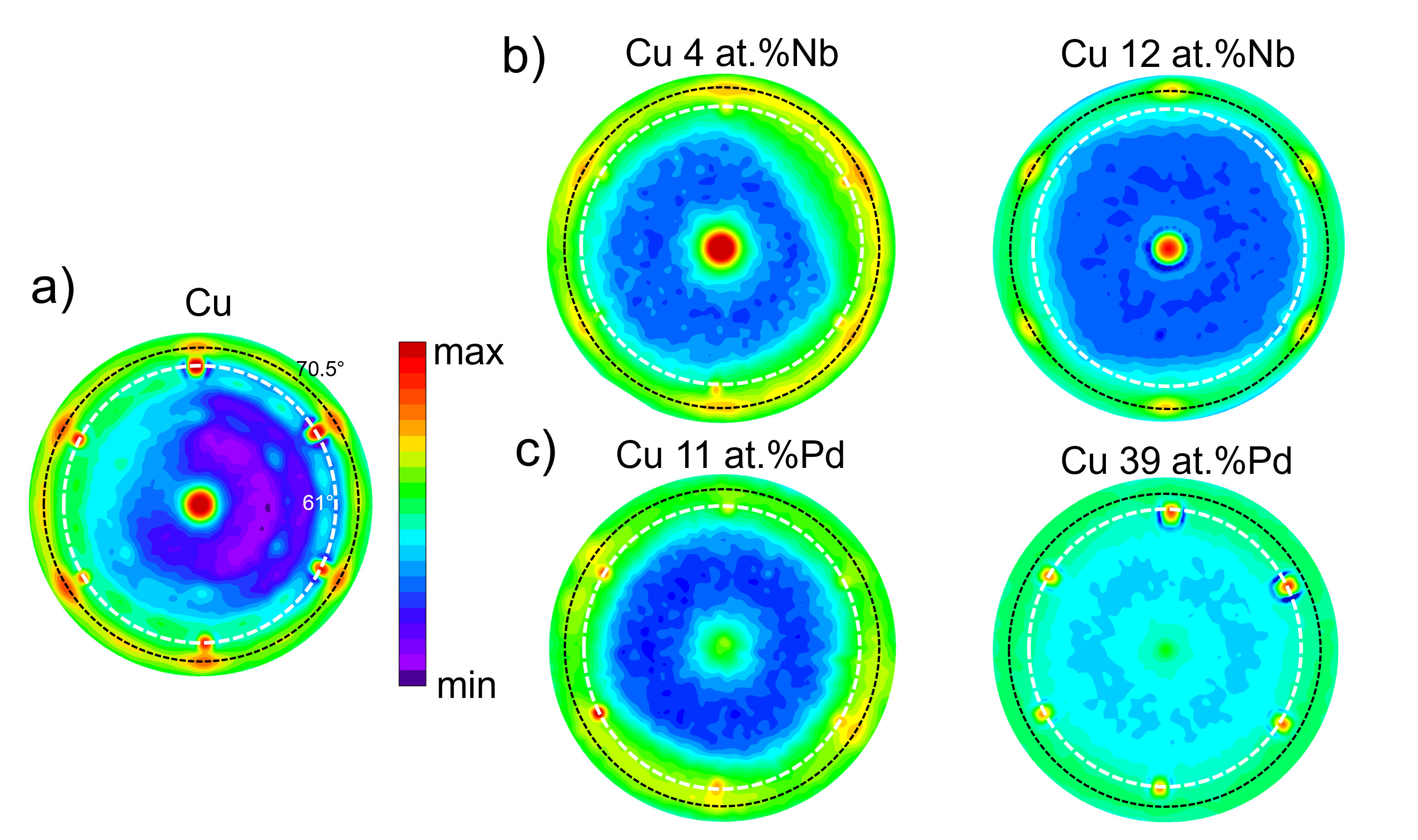}
\end{figure}

In figure \ref{pole}, pole figures of the Cu\{111\} family of planes are shown for selected samples compared to the pure Cu ( figure \ref{pole} a). Each reported pole figure has been acquired around the $2\theta$ position reported in Figs. \ref{CuNbfit} and \ref{CuPdfit} below. Six sharp spots appear at tilt of 62$^\circ$ (indicated in Fig.\ref{pole} with a white dashed concentric cicle) to the (2 -1 -1 3) substrate reflection which has 6-fold symmetry and which has a Bragg angle close to 43$^\circ$. These substrate spots become weaker when departing from the Bragg condition, as in the case of Cu 12 at.\% Nb. The pole figures of selected representative samples having different Nb and Pd concentrations, show similar features in term of out-of-plane and in-plane orientations. The crystallographic orientation relationship found in plane and out of plane of Cu films is: Cu\{111\}[0 0 1]//Al$_{2}$O$_{3}$\{0001\}[10-10]. In particular, the expected position corresponding to a [111] preferential orientation is intensity at a tilt angle of 70.5$^\circ$, indicate as a black dashed circular ring in all the panels of Fig.\ref{pole}. For pure Cu, in Fig. \ref{pole} a), other peaks very weak in intensity, are visible at 57 and 35$^\circ$ tilt angle, denoting a slight polycristallinity present in the film which is however, highly textured along the [111] direction. In the presence of Nb and Pd in Figs.\ref{pole} b) and c), respectively, the in-plane texture becomes less pronounced until the disappearance in Cu 39 at.\% Pd. 

\subsection{XRD experimental data analysis}
\subsection*{Cu-Nb}
\begin{figure}
\caption{XRD diffractograms around Cu (111) reflection of Cu films with different at.\% Nb concentrations. The solid lines are the fit using the substitutional version of the model presented in the text. The diffraction curves are in linear scale and have been vertically shifted for clarity.}
\includegraphics[width=1.0\textwidth]{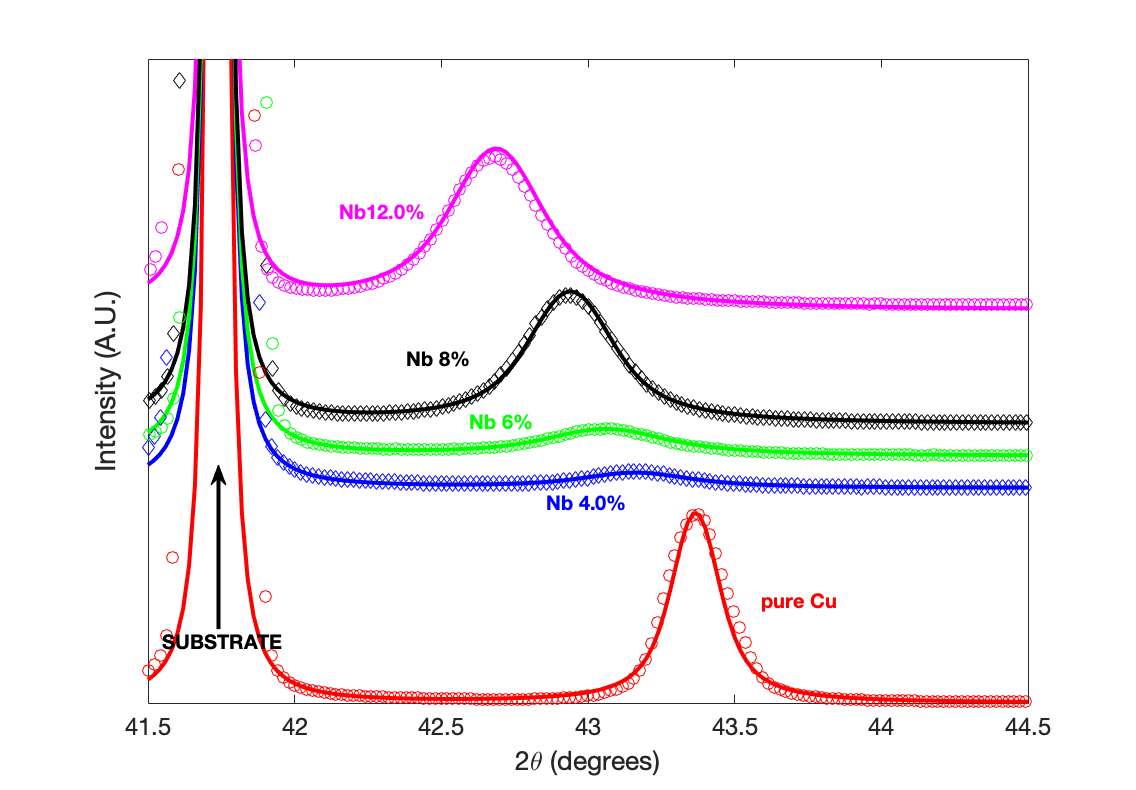}\label{CuNbfit}
\end{figure}

Surprisingly, the interstitial scenario we initially chose for the Nb/Cu system yields Nb concentrations that are consistently lower by a factor of two compared to those obtained from XRF measurements across the entire sample set. This result makes the interstitial model very unlikely to accurately describe the Cu–Nb system. In contrast, adjusting the data with the substitutional version, extracted Nb concentrations show an excellent agreement with the XRF values. This result, without any doubt, is revealing the substitutional nature of Nb doping of Cu for our room temperature co-evaporated samples. 
XRD $\theta$-2$\theta$ scans performed on Cu/Nb co-sputtered samples with 4 different Nb content, together with a pure Cu film are presented in figure \ref{CuNbfit}. 
The fitting curves were obtained with the substitutional version of the model presented in the section 2.1.

 \begin{table}
 \begin{center}
\caption{Parameters extracted from XRD analysis. From left to right: Nb concentration, normalized maximum of the (111) peak intensity with the corresponding diffraction angle, the resulting apparent interplanar distance, the effective grain size along the normal to the film (in \# of UC) with its standard deviation, the standard deviation of the internal positional disorder and the strain of the Cu matrix along the growth direction.}
\begin{tabular}{crcccccc}      
Nb at.\% & $ I^{norm}_{max}$ & 2 $\theta_{max} (^0)$ & $d_{AP}$ (Å) & N & $\Sigma$ & $\sigma$ (Å) & $\epsilon_z$(\%) \\
\hline
0 & 100.00   & 43.38 & 2.084 & 253 & 60 & 0.040 & -0.11\\
4 & 8.35  & 43.16 & 2.094 & 174 & 30 & 0.060 & -0.11\\
6 & 12.93 & 43.06 & 2.099 & 174 & 30 & 0.060 & -0.11\\
8 & 66.13 & 42.94 & 2.105 & 194 & 30 & 0.010 & -0.07\\
12 & 77.57& 42.68 & 2.117 & 233 & 30 & 0.010 &  0.04\\

\end{tabular}
\end{center}
\end{table}
\noindent In the first three columns of table 2 we report, for each Nb concentration  (at.\%), the normalized maximum of the (111) peak intensity ($ I^{norm}_{max}$), the corresponding diffraction angle (2$\theta_{max}$) and the resulting apparent interplanar distance ($d_{AP}$). In the last three columns, we show the model's parameters resulting from the fits:  the effective grain size along the normal to the film $N$ corrected from instrumental peak broadening (in \# of UC), its standard deviation $\Sigma$, the intrinsic Gaussian disorder $\sigma$ and the Cu matrix strain $\epsilon$ along the normal direction. A unique value ($\delta=0.256$ {\AA}) for the intercalation- induced local displacement,
was used to fit the data for all the compositions. The $\delta$ values are composition-independent and characterize the local distortion. They are intrinsic to the guest atom (Nb or Pd) when it is embedded in the Cu lattice. 
\begin{figure}
\caption{RCs analysis of Cu-Nb samples for different Nb contents (in at.\%) including the pure Cu film. The abscissa is the in-plane diffraction wave-vector. Solid lines are  Lorentzian fits. The sharp peak appearing in the center, is the tail of the (0006) peak of the sapphire. The RCs are vertically shifted for clarity.}
\includegraphics[width=0.8\textwidth]{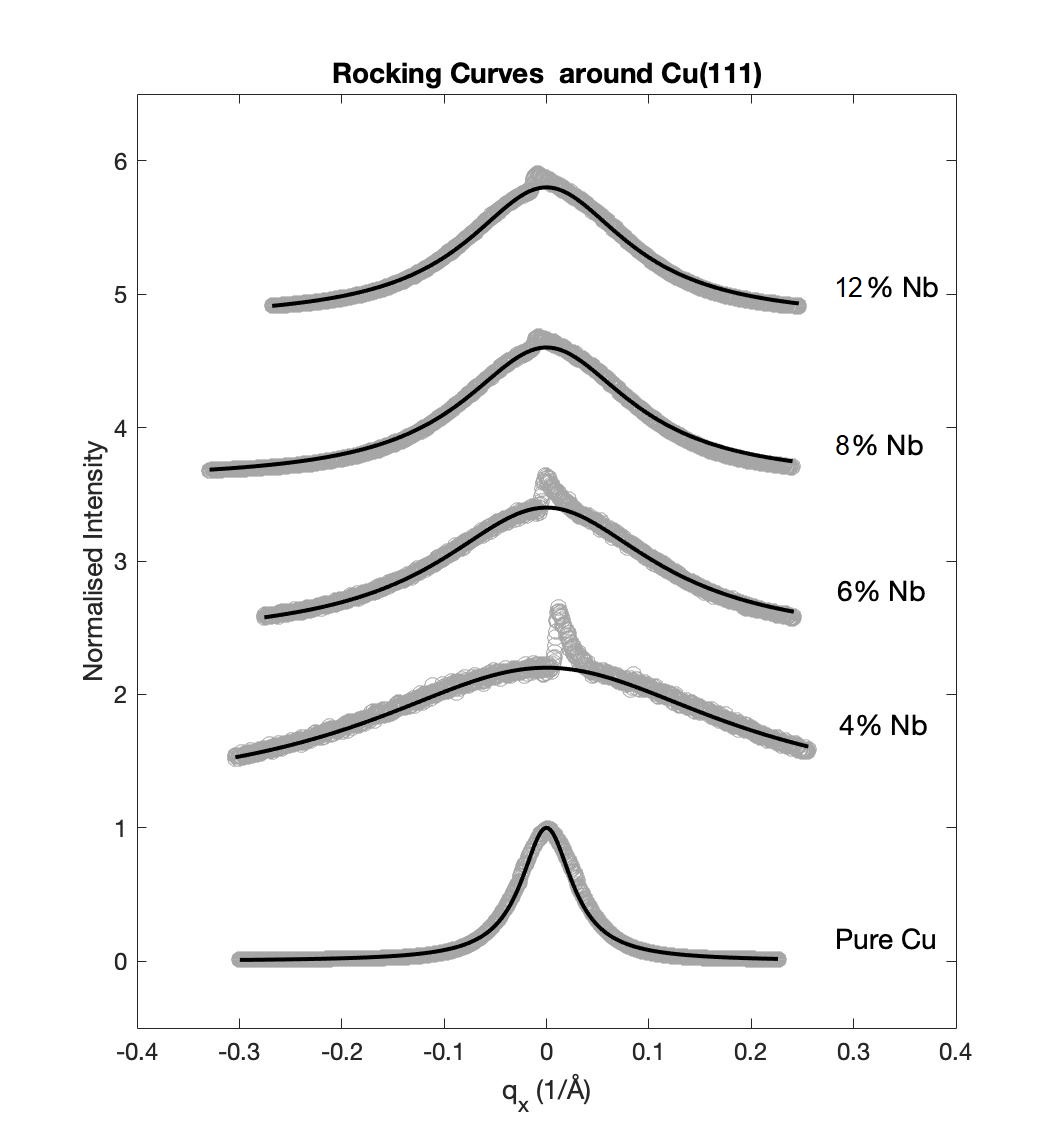}\label{RCCuNb}
\end{figure}

Rocking curves (RC) measurements have been performed  around each intensity maximum of the diffraction patterns of figure \ref{CuNbfit}. They are presented in figure \ref{RCCuNb} together with their corresponding Lorentzian fits. We recall that, for highly textured films, RCs are the Fourier transform of the in-plane electron density-density correlation function. Indeed, when the Gaussian polar distribution is narrow, the RCs are dominated by the exponential decay of the correlation function, whose Fourier transform (the diffracted intensity) is a Lorentzian. The typical decay distance, the crystal coherence length, identified with the average effective in-plane grain radius, is then proportional to the inverse of the FWHM of the RC. In Table 3 we list the values of the extracted  in-plane crystal coherence length   $\xi_x$ as a function of the Nb content. We list as well in the same table the out of plane values of effective grain sizes $\xi_z$ and elastic strain $\epsilon_z$ obtained previously from the $\theta-2\theta$ scans.
\\
\begin{table}
 \begin{center}
\caption{Nb content in at.\%, in plane crystal coherence length $\xi_x$ extracted from RC analysis,  out of plane crystal coherence length $\xi_z$ and out of plane strain for Cu/Nb co-sputtered thin films. The two last values previously obtained from $\theta-2\theta$ scans, have been reported from table 2 for convenience.}

\begin{tabular}{cccc}      
Nb at.\%& $\xi_z$ (\AA)& $\xi_x$ (\AA)& $\epsilon_z$(\%) \\
\hline

0 & 527 & 66 & -0.11  \\
4 &  361 & 9 & -0.11 \\
6 & 361 & 16 & -0.11 \\
8&403 & 20  & -0.07 \\
12& 485 & 21 & 0.04 \\

\end{tabular}
\end{center}
\end{table}
The evolution of the maximum intensity in the XRD diffractograms (Figure \ref{CuNbfit}) is somewhat unexpected. The intensity drops drastically going from the pure Cu layer to the sample doped with 4 at.\% Nb, then begins to increase monotonically gaining an order of magnitude for the sample with 10 at.\% Nb. This evolution seems in some way linked to the mesoscopic structure, in particular to the effective size of the grains in the plane. Indeed, the observed dramatic drop of in-plane coherence length from the pure Cu sample to the 4 at.\% Nb sample and its subsequent partial recovery for higher Nb contents appears to follow the same trend as the XRD intensity. 
\begin{figure}
\caption{Log-log plot of maximum intensity  $\it vs$ $\xi_x$, in-plane crystal coherence length for Cu/Nb co-sputtered films.}
\includegraphics[width=0.8\textwidth]{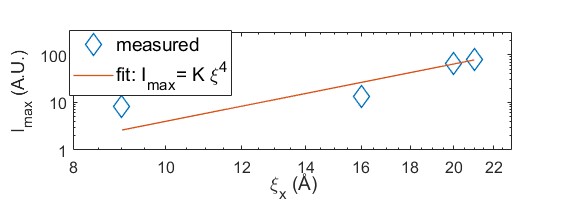}\label{Nbloglog}
\end{figure}
By normalizing the intensity with respect to the pure Cu and plotting it against the in-plane crystal coherence length we observe, in a log-log plot, a power law behavior consistent with $\xi_x^4$ as depicted in figure \ref{Nbloglog}. Co-sputtered Cu/Nb films present aspect ratios $\xi_x/\xi_z$ well below unity, characteristic of needle-like grain structures aligned along the film normal.  This implies, for an incoherent X-ray source (standard Cu-tube) and the relatively small angle of incidence ($\theta$ $\approx$ 21.5$^\circ$), a  power law of order 4 for the intensity. This scaling can be explained by the needle-like, highly textured nature of the investigated samples. Here, the intensity is not determined by the total crystallite volume. Instead, the diffracted signal is limited by the intersections of the X-ray beam with consecutive vertical grain boundaries: at a fixed Bragg angle $\theta$, the diffracted amplitude, which is proportional to number of coherently diffracting (111) planes, scales as $\xi_x \tan(\theta)$. The contribution to the intensity will thus scale as $\xi_x^2$. The intensity will also be proportional to the planar area of the crystallite, i.e. $\xi_x^2$. These two squared contributions finally confer a $\xi_x^4$ dependence to the measured intensity. In the opposite limit of planar-flake-like structures, the number of intersected planes would be invariant resulting just in a $\xi_x^2$ variation for the intensity.


\subsection*{Cu-Pd}
\begin{figure}
\caption{XRD diffractograms around Cu (111) reflection of Cu films with different Pd at.\% concentrations. The solid lines are the fit using the model presented in the text. The diffraction curves are in linear scale and have been vertically shifted for clarity. When indicated, they also have been multiplied by 20.}
\includegraphics[width=1.0\textwidth]{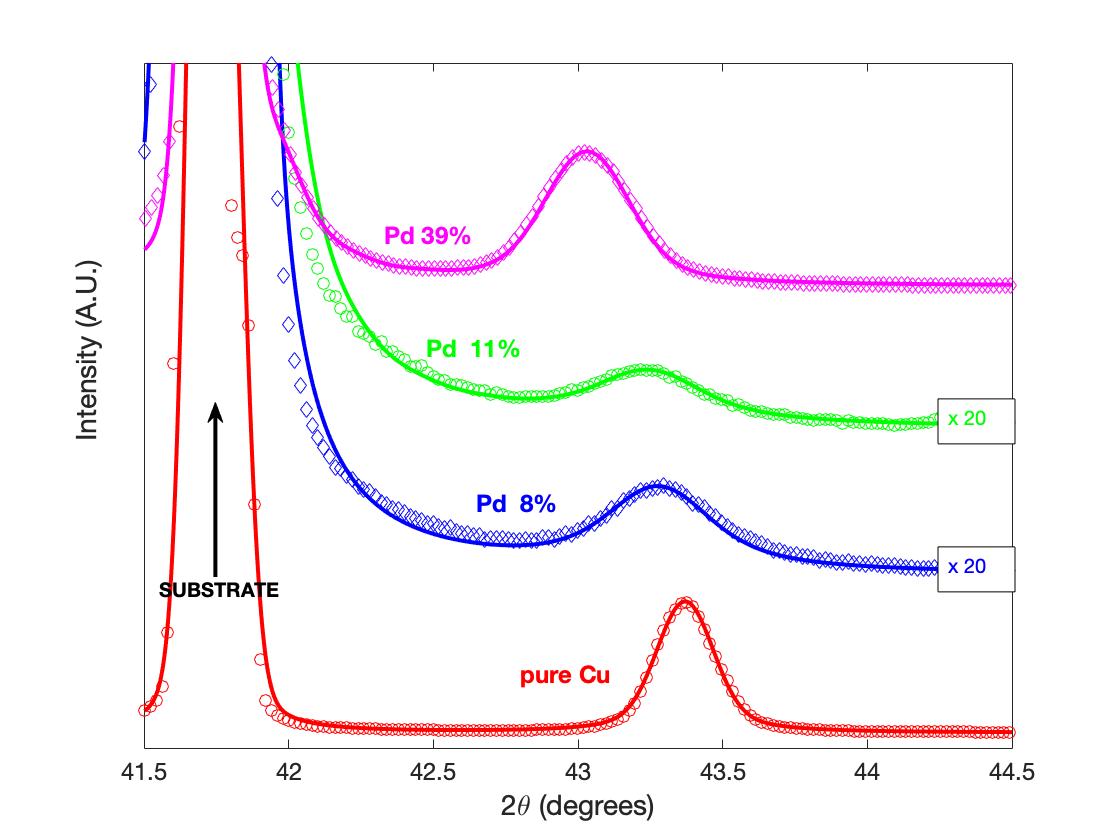}\label{CuPdfit}
\end{figure}

\begin{figure}
\caption{a) Fit of the XRD diffractogram for nominal Cu 39 at.\% Pd, and its decomposition  including the contribution of  (111) CuPd clusters. b) pole figure around the (111) position of the CuPd crystallites.}
\includegraphics[width=0.9\textwidth]{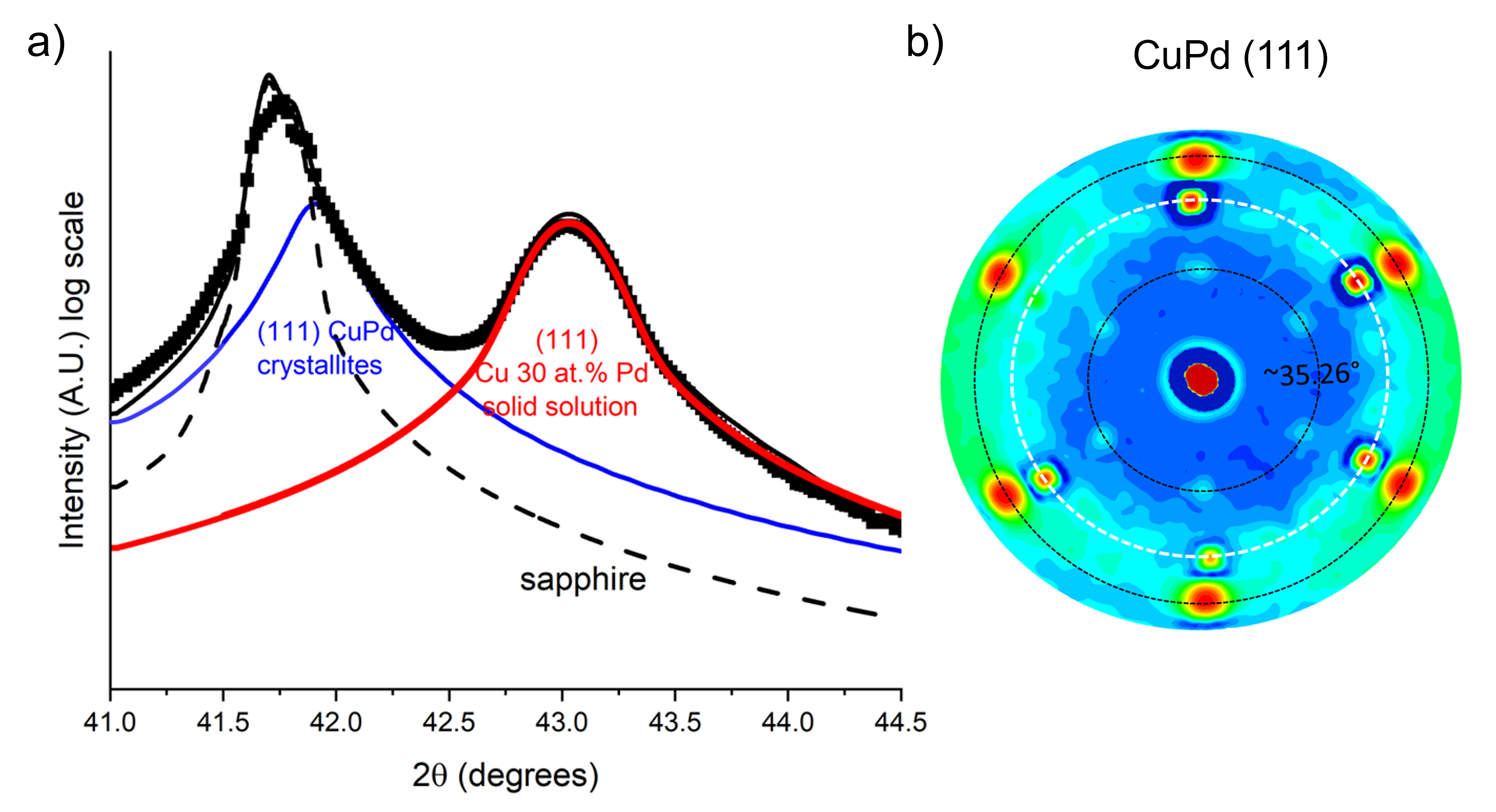}\label{CuPdphasesep}
\end{figure}

\begin{figure}
\caption{RC analysis for the Cu-Pd samples (numbers are in at.\%). The sharp peak
appearing in the center, is the tail of the (0006) peak of the sapphire. The RCs have been vertically shifted for clarity.}
\includegraphics[width=1.0\textwidth]{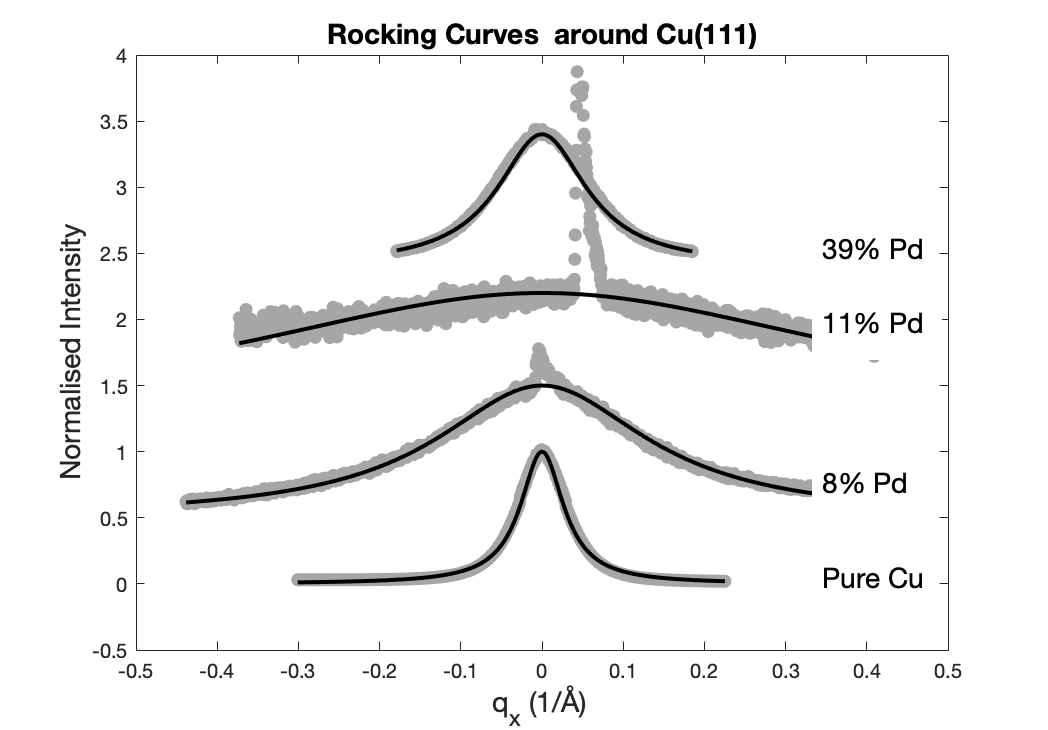}\label{CuPdRCs}
\end{figure}

For the miscible Cu-Pd system we have directly chosen the substitutional variant, as discussed in section 2.1. As for the Nb case, for the Cu-Pd alloy we report in Table 4 the model parameters resulting from the fits, together with other useful quantities from XRD analysis. 
 \begin{table}
 \begin{center}
\caption{Parameters extracted from XRD analysis. From left to right:  Pd concentration, normalized maximum of the (111) peak intensity with the corresponding diffraction angle, the resulting apparent interplanar distance, the effective grain size along the normal to the film (in \# of UC) with its standard deviation, the standard deviation of the internal positional disorder and the strain of the Cu matrix along the growth direction.}
\begin{tabular}{cccccccc}      
Pd at.\% & $ I^{norm}_{max}$ & 2 $\theta_{max} (^0)$ & $d_{AP}$ (Å) & N & $\Sigma$ & $\sigma$ (Å) & $\epsilon_z$(\%) \\
\hline
  0 & 100.0  & 43.38 & 2.084 & 253 & 60 & 0.04 & -0.11\\
  8 &  3.30  & 43.28 & 2.089 & 55 & 30 & 0.07 & -0.12\\
 11 & 2.10 & 43.23 &  2.091 & 68 & 30 & 0.08 & -0.08\\
39 & 97.0 & 43.03 & 2.100 & 106 & 20 & 0.03 & -0.12\\
\end{tabular}
\end{center}
\end{table}
\noindent 
The  unique value for the intercalation induced displacement used to fit the data for all the compositions was $\delta = 0.125$ {\AA}. In the last row of Table 4, the Pd content indicated as 39 at.\% is the nominal one. From the model adjustment, the actual concentration of Pd within the Cu-Pd solid solution is 30 at.\%, the remaining Pd is forming the CuPd crystalline phase. The fraction of the crystalline phase can be deduced by combining the measured XRF global Pd content (39 at.\%) and the Pd concentration within the solid solution given by the random intercalation model adjustment (30 at.\%). Simple algebra yields  45 at.\% for the stoichiometric crystalline phase of CuPd. The contribution of stoichiometric CuPd compound to the XRD pattern is drowned in the tail of the sapphire substrate peak. However, the decomposition shown in fig. \ref{CuPdphasesep} a) highlights the presence of this peak with a weight similar to the one of the solid solution, confirming our previous determination. The pole figure performed around the CuPd (111) Bragg condition is shown in Fig. \ref{CuPdphasesep} b). The FCC structure and the [111] preferential orientation in this phase are evidenced by the 6 bright spots measured at a tilt angle of 70.5$^\circ$ (dashed black line), exactly like in the case of pure Cu. Other 6 weaker spots are present at 35$^\circ$ indicating the presence of polycristallinity and other orientations. 
In figure \ref{CuPdRCs} we present the RC's measured for three different Pd compositions. The curve in the top part of the figure, while is labeled with 39 at.\% Pd for its nominal concentration, corresponds to the RC taken around the peak at $2\theta =43^\circ$ which actually corresponds to the solid solution with 30 at.\% Pd.
\noindent As indicated in the preceding subsection, we extract the in-plane crystal coherence length  from the Lorentzian fit of the RCs, and we list them in Table 5, together with the out of plane crystal coherence length   and the strain along the normal to the film.\\

\begin{table}
 \begin{center}
\caption{Pd content in at.\%, in plane crystal coherence length $\xi_x$ extracted from RC analysis,  out of plane crystal coherence length $\xi_z$ and out of plane strain for Cu/Pd co-sputtered thin films. The two last values previously obtained from $\theta-2\theta$ scans, have been reported from table 4 for convenience.}

\begin{tabular}{cccc}      
Pd at\%.& $\xi_z$ (\AA)& $\xi_x$ (\AA)& $\epsilon_z$(\%) \\
\hline

0 & 526 & 66 & -0.11  \\
8 &  115 & 16 & -0.12 \\
11 & 142 & 5 & -0.08 \\
39 &221 &38 & -0.12 \\

\end{tabular}
\end{center}
\end{table}

\begin{figure}
\caption{Log-log plot of maximum intensity  $\it vs$ in-plane crystal coherence length for Cu/Pd co evaporated films.}
\includegraphics[width=0.8\textwidth]{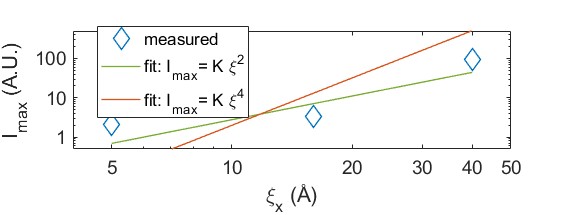}\label{Pdloglog}
\end{figure}

\noindent The XRD intensity and the RC for Cu 11 at.\% Pd is apparently out of the trend observed previously in the case of Nb. However, the simultaneous abrupt drop off of intensity and in-plane $\xi_x$ at low Pd content and the partial recuperation of both quantities at higher Pd concentrations, persists. 
To complete the analysis, as for the Nb case, in fig. \ref{Pdloglog} we present a log-log plot of the normalized intensity as a function of the in-plane coherence length to find the corresponding power law. 
\noindent In the case of Pd, the power law is not clear since, at first glance, a $\xi^2_x$ scaling seems to be more adequate. However, this scaling is expected for much higher aspect ratios. Excluding the sample at 11\% Pd (the one with smaller in-plane crystal coherence length that appears to be out of the main trend) the intensity of the two remaining samples is in good agreement with the scaling in $\xi^4_x$. 
\subsection{Comparison of Cu-Pd and Cu-Nb}
The evolution of XRD and mesoscopic structure for the two series of samples, Cu-Nb and Cu-Pd, is complex. Indeed, the intercalation of Nb (Pd) within the Cu matrix introduces, at first, structural defects resulting in smaller crystallite sizes and strongly weakening of the XRD intensity of the Cu (111) reflection. This effect has been already observed in co-sputtered Al-Si alloys \cite{CANCELLIERI2017120} where the effect of
Si introduction as low as 4 at.\% in the PVD grown layers is dramatic for the preferential
[111] orientation in Al. 
Then, as the amount of the secondary metal content increases to 12 at.\% for Nb and to 39 at.\% for Pd, the  in-plane and out of plane crystal coherent length (crystallite effective  size) increases together with the XRD intensity, partially recovering to the level of pure Cu layers. As already investigated in Ref. \cite{Petrov}, the structure evolution during nucleation of co-sputtered thin films is complex and governed by many factors: surface and bulk diffusion/mobility, solubility of the co-sputtered materials, energetic of particles, grain boundary and surface segregation. The interplay of all these factors affect the film formation process, limiting or favoring grain coarsening during coalescence and even interrupting epitaxial growth of individual crystallites. Immiscible elements co-deposited
using physical vapor deposition (PVD) can self-organize into
phase-separated regions with chemically sharp interfaces \cite{XUE2019162, Derby02012019}. Additionally, substitutional incorporation of Nb into the Cu lattice is also consistent with  synthesis of thin films under out-of-equilibrium conditions \cite{MA2005413}, i.e. magnetron sputtering. Techniques such as rapid quenching, thin film deposition, and severe plastic deformation have been shown to stabilize substitutional Nb atoms in the Cu lattice (CuNb alloys), even though the equilibrium solubility of Nb in Cu is extremely low. These methods promote metastable solid solutions by kinetically suppressing phase separation, allowing Nb to occupy Cu lattice sites rather than forming intermetallic compounds or segregating at grain boundaries. Our findings, which indicate substitutional Nb doping are therefore in line with these observations from out-of-equilibrium synthesis routes. In perfect agreement  with the substitutional scenario, the larger distortion parameter $\delta$ obtained for Cu–Nb compared to Cu–Pd (0.256 \AA~versus 0.125 \AA) is consistent with the relative atomic radii of Nb and Pd with respect to Cu: since Nb is approximately twice the size mismatch of Pd, its substitution in the Cu lattice is expected to induce a stronger local distortion, justifying the doubled $\delta$ parameter in the random intercalation model.
Although the complete understanding of XRD intensity recovery with increasing Pd and Nb concentration requires further analysis, it is presumably related to the hierarchical organization \cite{Powers} when condensation occurs, of the species simultaneously present in the magnetron plasma.  Moreover, there is a clear correlation between the crystallite structure and the measured XRD intensities. Both series have aspect ratios $\xi_x / \xi_z$ well below unity, characteristic of needle grain structures along the film normal and their intensity scales as $\xi_x^4$. Therefore, we can explain the singular intensity evolution by the needle like structure induced  by the introduction of extra elements (Nb and Pd) in Cu/Nb and Cu/Pd co-sputtered films.

\begin{figure}
\caption{Evolution of the d$_{111}$ spacing as a function of Nb and Pd content, compared to Vegard's law for Pd. The error bars are estimated from the uncertainty in the peak position of Figs. 2 and 5 from a Pseudo-Voigt fitting.}
\includegraphics[width=1.0\textwidth]{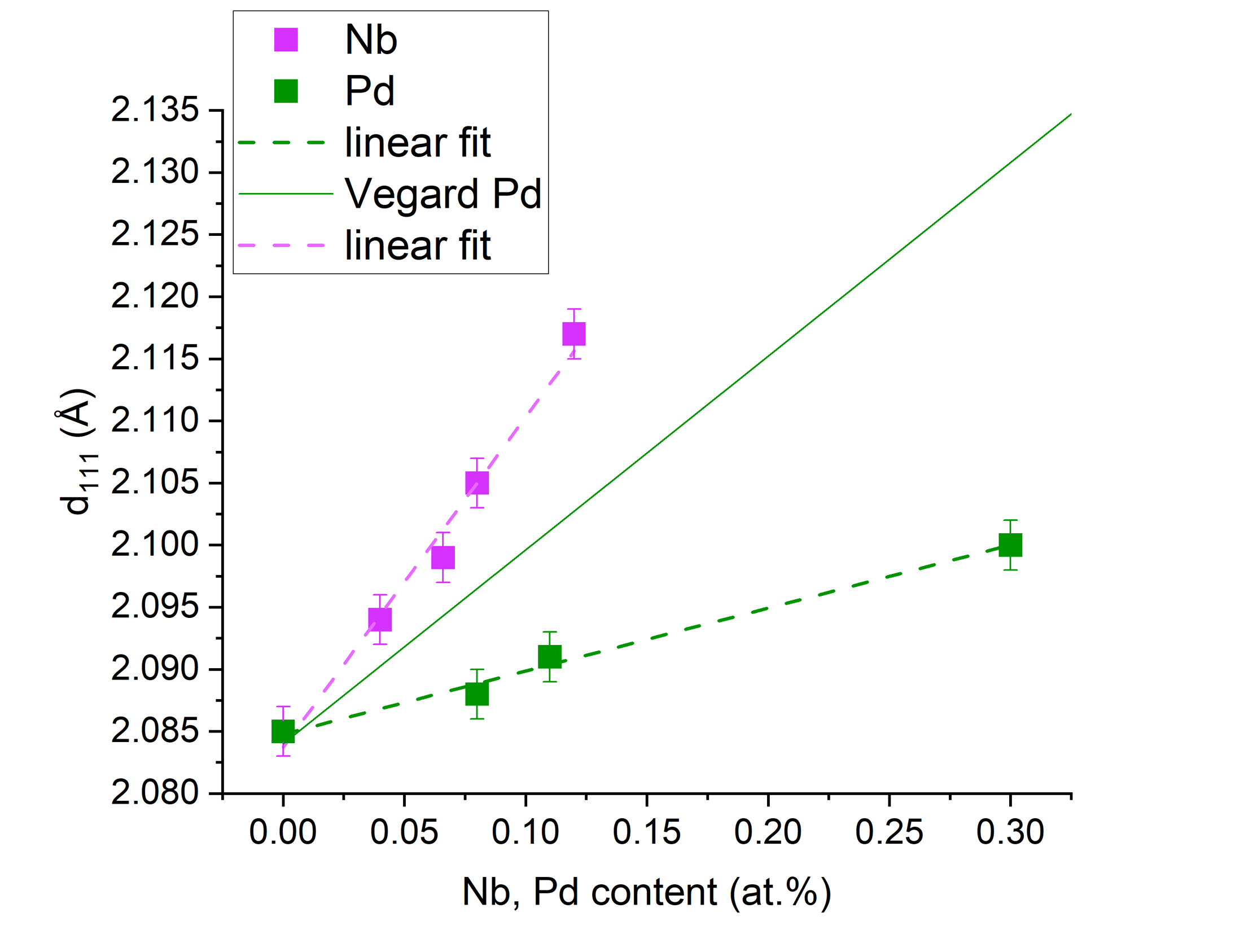}\label{veegards}
\end{figure}

Another characteristic that should be observed among our results is the linear behavior of the apparent interplanar distance (calculated from peak position) in the Cu matrix along the [111] direction as a function of the Nb or Pd content $p = \alpha$. This dependence is illustrated in Figure \ref{veegards}. 

\noindent Although this linearity is reminiscent of Vegard's law, we should emphasize that it is meaningless in our case. Indeed, Vegard's law is normally defined  near thermodynamic equilibrium because metastable phases, strain, or defects can cause deviations from the ideal linear trend in solid solution alloys. Thermodynamically, Nb atom cannot form a solid solution with Cu as they are immiscible, although, in non-equilibrium conditions e.g., magnetron co-sputtering thin films deposition, they can form metastable solid solution. Even in this case, Vegard's law would predict a negative slope of the line based on the difference in lattice parameters between Cu and Nb, which is the opposite of what it is experimentally measured in Fig. \ref{veegards}. In the case of Pd, the distribution of the guest atoms among the host lattice sites is not governed by thermal equilibrium, preventing the system to find the global average lattice constant predicted by Vegard's law. In the case of Cu/Pd samples, the slope of the $d_{111}$ $vs$ Pd content is 3 times smaller than expected from Vegard's law. Moreover, the observed linearity is  exactly predicted in both cases by the random intercalation model and reads simply as : $d_{111}(\alpha)=d^0_{111}+ \alpha \delta$, where $d^0_{111}$ is the planar spacing for pure Cu.

At room temperature growth, Nb ad-atoms are less mobile than Cu ad-atoms due to much higher melting point (and so homologous temperature) of Nb. Sputtered species have very high kinetic energies \cite{Rossnagel} and during film formation these species are very rapidly quenched to the substrate temperature.  At such non equilibrium conditions, the micro-structural evolution is governed by kinetics rather than thermodynamics. At low Nb or Pd content, Nb / Pd atoms remain ``frozen" on non-equilibrium surface sites at the film growth front. This Nb/Pd ad-atoms then act as nucleation sites for more mobile Cu leading to repeated nucleation of Cu islands. This gives rise to the interruption of local epitaxy and grain refinement. Limited availability of Nb/Pd combined with higher barrier to diffuse over significant distances can result in small precipitate-clusters usually with the sizes below detection limit of XRD. Depending on Nb/Pd concentration and substrate temperature, a more complex microstructure can evolve. In the next section, the thin film microstructure is analyzed in function of the Nb and Pd content.
 
     \subsection{SEM and STEM analysis}
\begin{figure}
\caption{SEM planar views of a) pure Cu. b) Cu 4 at.\% Nb, c) 10 at.\% Nb, d) Cu 8 at.\% Pd and Cu 39 at.\% Pd. The different in plane dimensions of the grains is evident.}
\includegraphics[width=1.0\textwidth]{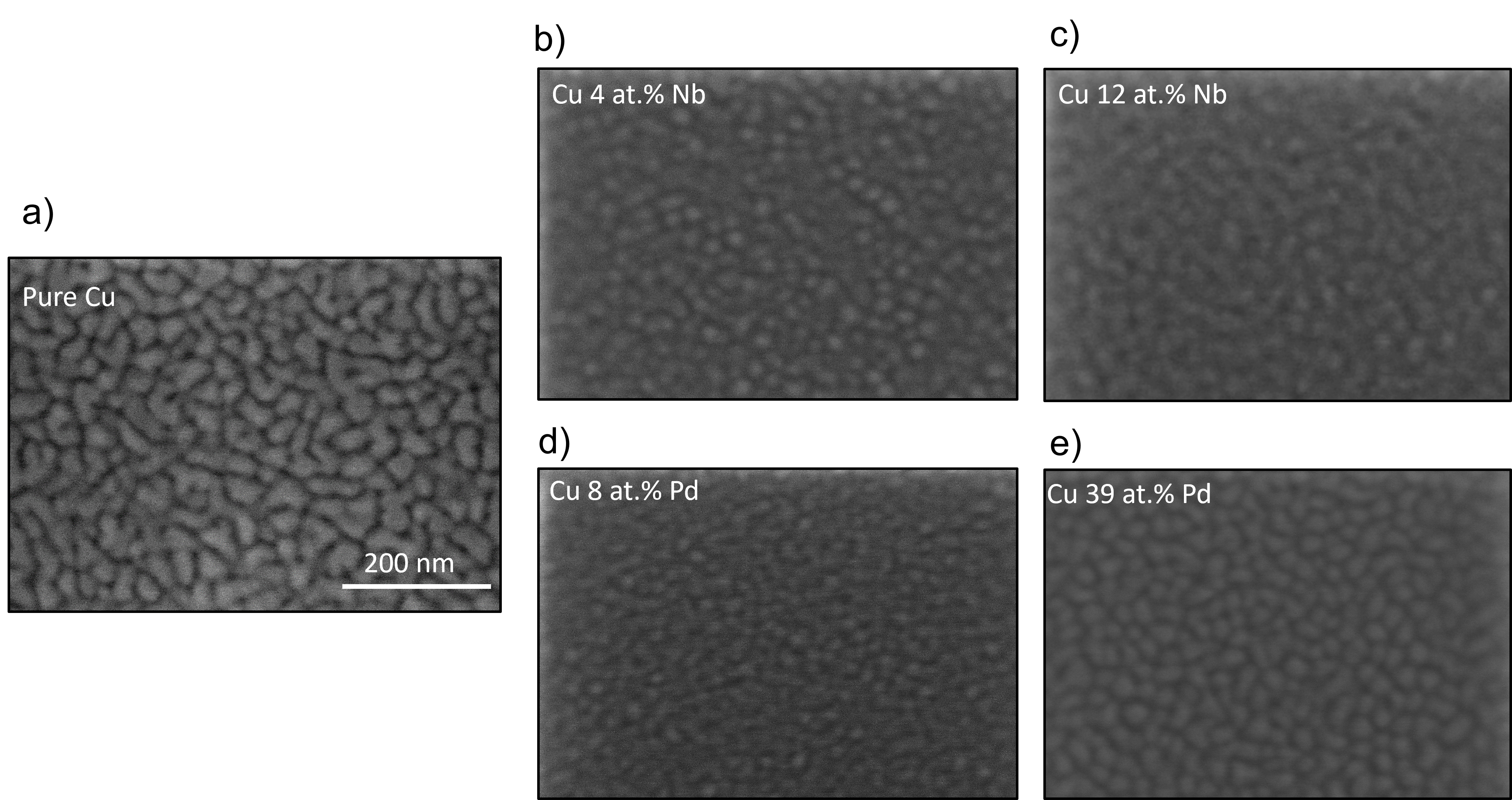}\label{SEM}
\end{figure}

 In figure \ref{SEM}, surface images of the thin films of Cu-Pd and Cu-Nb with the most extreme compositions, are shown compared to pure Cu (Fig.~\ref{SEM}a)).
The different grain dimensions are evident: 
Figs.~\ref{SEM}b) and d) display an average in plane grain size of 5-10 nm while Figs.~\ref{SEM} c) and d) the grains are between 20-30 nm, similarly to pure Cu.
This is in line with the results shown in the previous section on the RC analysis. Although the crystal coherence lengths and grain size from microscopy analysis may have different magnitude as they measure different quantities, a correlation is expected. More specifically, the crystal coherence lengths extracted from XRD are not determined by the density of well defined structural defects as grain boundaries, like for usual single composition materials. In the present case, the cumulative continuous loss of phase coherence along the beam path due to random guest intercalation is a more subtle mechanism. Indeed, it affects the correlation function on a shorter scale than abrupt defects related to grain formation. We can have, thus, well textured grains showing a crystal coherence length from diffraction much shorter than the grain size visualized by electron microscopy.
Importantly, the trend of the $\xi_x$ extracted from the RC analysis in Figs.\ref{Nbloglog} and \ref{Pdloglog} follows the one found for the in plane grain size from the SEM pictures: the grain size increases again with increasing the Pd or Nb content.

\begin{figure}
\caption{EDX surface analysis in a) and STEM cross section with EDX elemental mapping in b) for a Cu 4 at.\% Nb film}
\includegraphics[width=1.0\textwidth]{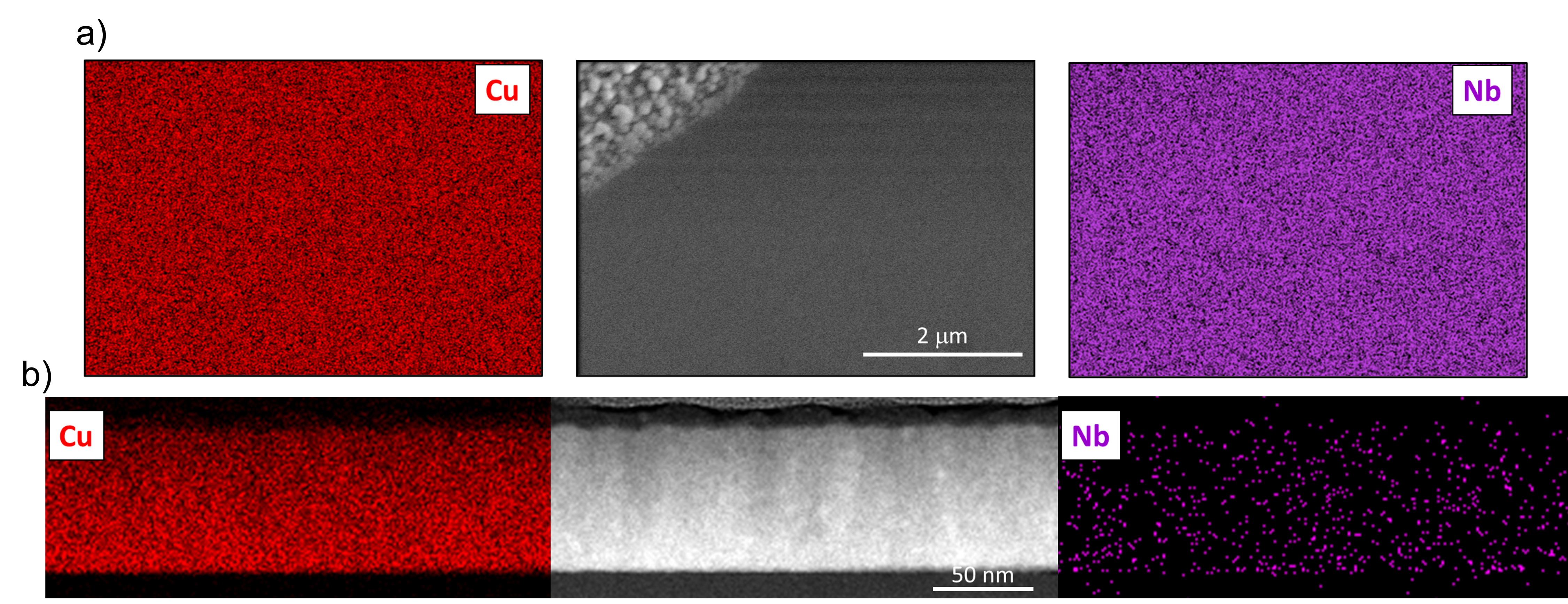}\label{STEMCuNb}
\end{figure}

\begin{figure}
\caption{EDX surface analysis in a) and STEM cross section with EDX elemental mapping in b) for a Cu 11 at.\% Pd film}
\includegraphics[width=1.0\textwidth]{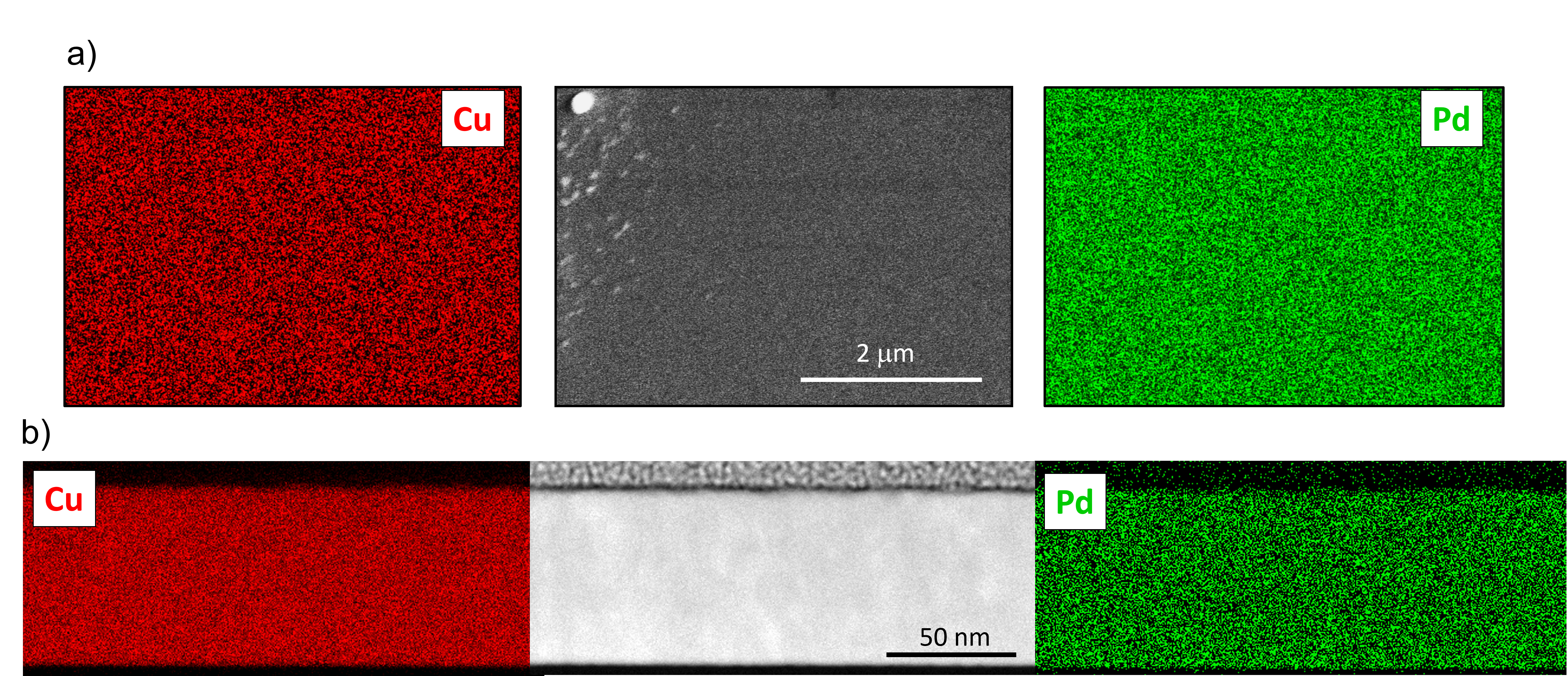}\label{STEMCuPd}
\end{figure}

\begin{figure}
\caption{EDX surface elemental mapping for a Cu 39 at.\% Pd film.}
\includegraphics[width=1.0\textwidth]{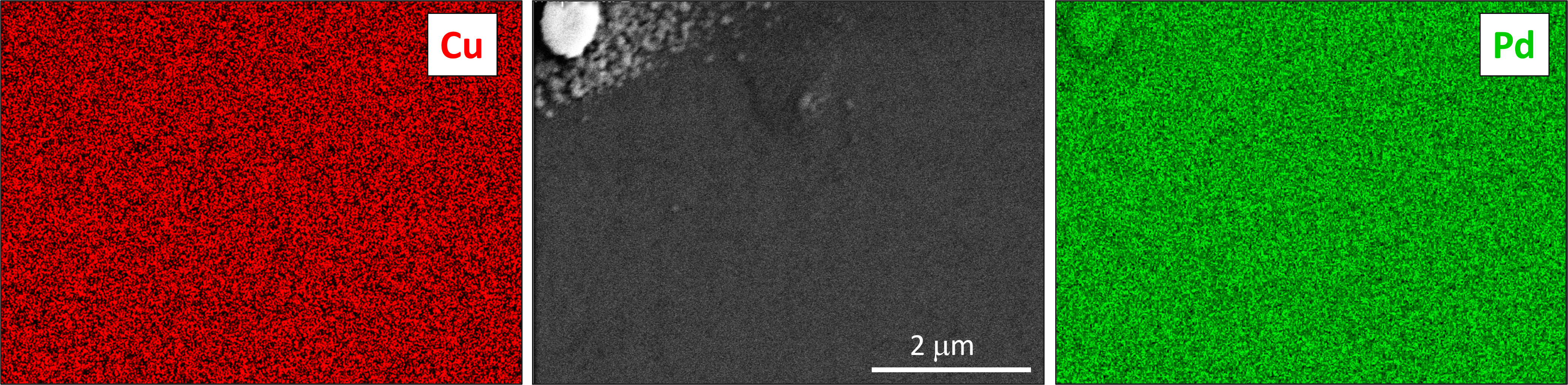}\label{STEMCuPd40}
\end{figure}

To examine the elemental distribution both in-plane and out-of-plane, we performed EDX analysis of Cu, Nb, and Pd.
The results for Cu 4 at.\% Nb are shown in Fig.~\ref{STEMCuNb}. The images, both in-plane and in cross-section, indicate a homogeneous distribution of Nb within the Cu matrix. Specifically, in Fig.~\ref{STEMCuNb}(a), a surface region containing some agglomerated particles was selected; however, the EDX analysis did not reveal Nb phase-separated grains.
The same analysis, conducted on the Cu 11 at.\% Pd sample, is shown in Fig.~\ref{STEMCuPd}.
In this case as well, Pd is homogeneously distributed within the Cu lattice, forming a solid solution with no clear evidence of phase separation or agglomeration. The analysis of the Cu 39 at.\% Pd sample shown in Fig.~\ref{STEMCuPd40}, displays also an homogeneous distribution of Pd and although a contrast is seen for a Pd and not for Cu elemental mapping, it is difficult to conclude that there is Pd accumulation from SEM analysis. The XRD fit results for this high Pd concentration, shown in Fig.~\ref{CuPdphasesep}, clearly shows the contribution of a stoichiometric CuPd compound alongside the Cu/Pd solid solution. Thus, our XRD fitting, which captures lattice parameter shifts and shoulder intensities, provides the most reliable evidence of Pd accumulation. However, these crystallites are probably smaller than the STEM resolution for elemental mapping.

\section{Conclusions}
In conclusion, this study reveals a complex relationship between elemental concentration and the structural properties of Cu/Nb and Cu/Pd co-sputtered thin films.  A random intercalation model considering both interstitial and substitutional incorporation of guest atoms was presented. Our analysis demonstrates that only the substitutional scenario accurately reproduces the XRD data for all Nb and Pd concentrations. The larger local displacement induced by Nb compared to Pd is fully consistent with the substitutional model and reflects the difference in atomic radii between Nb and Pd.  While low levels of Nb and Pd lead to reduced crystal coherence and lower XRD intensities, likely due to the limited mobility of the dopant atoms and their role as nucleation sites for Cu, higher second element concentrations result in a moderate increase of the in-plane crystal coherence lengths. Additionally, coherent domain structures show very low aspect ratios, giving them needle-like shapes with the long axis aligned along the normal to the films.
The immiscible Cu/Nb system, explored from 0 to   12 at.\% Nb does not show Nb clustering. Conversely, for the Cu/Pd system, the 39 at\% Pd film splits in a 30 at\% Pd solid solution and embedded crystalline clusters of the cubic CuPd compound. This observation is in accordance, for this concentration, with the published phase diagram \cite{POPOV2019204}.
The XRD-derived structural information are supported by the SEM and EDX analysis. The proposed model can be applied to other thin film systems of co-deposited species.

\appendix



\ack{Acknowledgements}

The authors thank Robin Bucher for his help with the SEM and EDX measurements. The authors acknowledge Lars P.H. Jeurgens and Bastian Rheingans for fruitful discussions. The authors thank Agnes Mill for FIB preparation.
D.A. wishes to thank the support of the Uruguayan  institution PEDECIBA and the SNI (Sistema Nacional de Investigadores- ANII).
\bibliographystyle{iucr}
\bibliography{iucr}




\end{document}